\documentclass{article}

\usepackage{aimc2025}
\usepackage{float}
\usepackage[utf8]{inputenc} 
\usepackage[T1]{fontenc}    
\usepackage{hyperref}       
\hypersetup{
  colorlinks   = true, 
  urlcolor     = blue, 
  linkcolor    = black, 
  citecolor   = black 
}
\usepackage{url}            
\usepackage{booktabs}       
\usepackage{amsfonts}       
\usepackage{nicefrac}       
\usepackage{microtype}      
\usepackage{graphicx}       
\usepackage{tabularx}
\usepackage{makecell}
\usepackage{censor}

\title{The Shape of Surprise: Structured Uncertainty and Co-Creativity in AI Music Tools}

\author{
  Eric Browne\thanks{\url{https://ericbrowne.com}} \\
  MTU Cork School of Music\\
  Cork, Ireland \\
  \texttt{\url{eric@ericbrowne.com}} \\
}

\begin{document}

\maketitle

\begin{abstract}
Randomness plays a pivotal yet paradoxical role in computational music creativity: it can spark novelty, but unchecked chance risks incoherence. This paper presents a thematic review of contemporary AI music systems, examining how designers incorporate randomness and uncertainty into creative practice. I draw on the concept of \textit{structured uncertainty} to analyse how stochastic processes are constrained within musical and interactive frameworks. Through a comparative analysis of six systems—Musika \citep{pasini2022musika}, MIDI-DDSP \citep{wu2021midi}, Melody RNN\footnote{\url{https://magenta.tensorflow.org/2016/07/15/lookback-rnn-attention-rnn/}}, RAVE \citep{caillon2021rave}, Wekinator \citep{fiebrink2010wekinator}, and Somax 2 \citep{borg2019somax}—we can identify recurring design patterns that support musical coherence, user control, and co-creativity. To my knowledge, this is the first thematic review examining randomness in AI music through structured uncertainty, offering practical insights for designers and artists aiming to support expressive, collaborative, or improvisational interactions.
\end{abstract}

\section{Introduction}

The interplay between determinacy and indeterminacy has long been a driver of musical creativity. Composers and system designers have leveraged randomness as a source of surprise and innovation, from early algorithmic compositions to modern AI-generated music. In mid-20th-century experiments, pioneers like Hiller \citep{hiller1979experimental} and Xenakis \citep{xenakis1992formalized} introduced controlled randomness – using chance operations guided by rules or probability distributions – to generate complex musical textures. This approach showed that randomness, when constrained appropriately, could yield musically interesting results beyond the reach of purely deterministic methods. The same principle carries into AI music: random variation helps generative models break out of predictable patterns, but musical expectations impose constraints. If an AI music system outputs notes or sounds completely at random with no regard for musical context, the result will likely be unsatisfying.

Rather than eliminating randomness, successful creative systems incorporate it within frameworks that preserve musical sense. One useful way to describe this is through the lens of \textit{structured uncertainty}—an approach that introduces unpredictability within meaningful boundaries to promote creative outcomes. While the term structured uncertainty is not universally established in music AI, related concepts appear in adjacent fields. \citet{beghetto2019structured} explicitly frames structured uncertainty as a creative affordance in education: \begin{quote}
    "The concept of structured uncertainty builds on the commonly agreed upon criteria necessary for judging something as creative. It also illustrates how slogans like “think outside the box” and similar conceptions that portray creativity as a form of unstructured originality are problematic." \cite{beghetto2019structured}
\end{quote} While Beghetto's introduction of the term is designed more specifically for organising structured learning experiences for creativity, I draw on the concept of \textit{structured uncertainty} to analyse how stochastic processes are constrained within musical and interactive frameworks. We can see practical examples of similar approaches successfully implemented across a range of creative disciplines. For example, \citet{ibert2018uncertainty} discuss how professionals dynamically navigate and structure uncertainty in creative domains. \citet{daikoku2021statistical} show that intermediate levels of uncertainty boost cognitive engagement and creative potential in musical contexts.

These perspectives suggest that designing generative algorithms with structured uncertainty—allowing for surprise while adhering to learned musical structures or user-defined constraints—can actively promote purposeful exploration without undermining coherence. For example, a model might use probability distributions learned from data to ensure its random choices reflect a certain style, or it might allow a human to steer random variations during the creative process. The challenge is to enable intentional unpredictability: providing the catalyst of chance operations while retaining influence over the high-level outcome. 

While my usage of ‘structured uncertainty’ is inspired by Beghetto’s framing in educational settings, I adapt the term as a conceptual lens for analysing AI music systems. The aim is not to directly apply his framework, but to repurpose its core idea—uncertainty bounded by structure—as a way to think through design tensions in musical tools.

In this paper, I present a thematic review of randomness in AI music systems, examining how different systems handle uncertainty in their generative processes. I identify key dimensions that characterise various approaches—such as whether randomness is introduced in an unstructured or structured manner, and the degree of human interaction involved. This is followed by a comparative analysis of six selected AI music systems that exemplify these approaches. While the primary aim is to trace common patterns in the design of structured uncertainty, I also reflect on how these strategies may shape systems’ potential for co-creative interaction, particularly in real-time or responsive contexts. By doing so, I point out not only the technical affordances of randomness, but also its role in enabling richer forms of human-AI musical collaboration.

\begin{figure}[ht]
  \centering
  \includegraphics[width=0.85\linewidth]{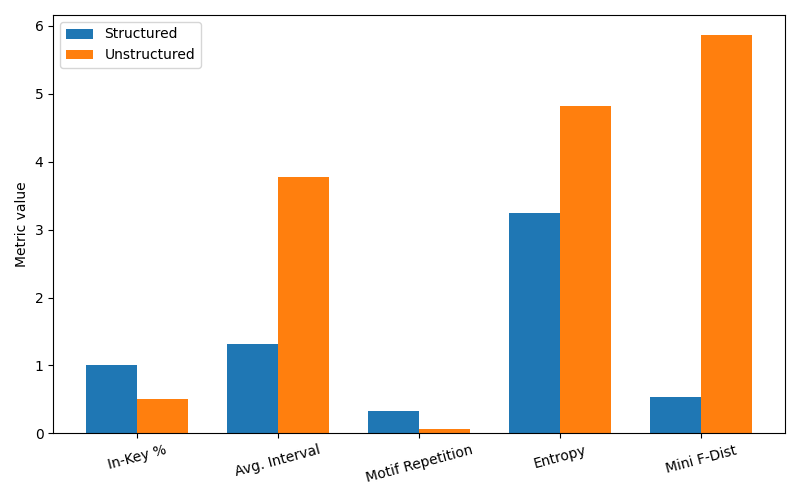}
  \caption{Effect of structural constraints on stochastic melody generation. 
  Thirty-two-note sequences are generated using either a first-order Markov model trained on “Twinkle Twinkle Little Star” (structured) or uniform random pitch selection (unstructured). Five illustrative metrics are shown: percentage of notes in key, average melodic interval size, motif repetition rate, transition entropy, and a toy Fréchet-style distance to the training melody, adapted from \citep{retkowski2024frechet}. Higher values are preferable for in-key percentage and motif repetition; lower values indicate smoother motion (interval size), greater predictability (entropy), and closer similarity to the original melody (Fréchet distance). While simplified and not definitive indicators of musical quality, these metrics demonstrate how even basic structural constraints significantly strengthen the coherence of stochastic outputs.}
  \label{fig:structured-randomness}
\end{figure}

\section{Background: Randomness and Uncertainty in Music AI}

\subsection{Randomness in Musical Creativity}

Chance operations have a rich history in music, exemplified by early algorithmic compositions that employed randomness deliberately. Iannis Xenakis notably used mathematical probability distributions to compose \textit{stochastic music}—textures unpredictable in detail but structured at an aggregate level \citep{xenakis1992formalized}. This historically established the idea that uncertainty, when structured, can serve as an intentional catalyst for creativity, aligning unpredictability with artistic coherence. \citet{borgo2005sync} explores how improvisational systems often rely on this dynamic interplay between order and indeterminacy, suggesting that creative emergence may depend on flexible responsiveness within complex musical environments.

\subsection{Randomness in Algorithmic and AI Systems}

As music generation algorithms evolved, randomness became a key component. Traditional algorithmic composers often employed random number generators within rule-based frameworks (e.g., rolling virtual dice to choose notes under certain constraints). In modern AI, machine learning models incorporate randomness in several ways – from stochastic optimisation during training to probabilistic sampling during generation. For instance, generative adversarial networks (GANs) \citep{donahue2018adversarial} and variational autoencoders (VAEs) \citep{roberts2017hierarchical} use random latent vectors to generate novel outputs, and recurrent neural networks (RNNs) \citep{eck2002first} can stochastically choose each subsequent note in a sequence. In all these cases, randomness is a tool to traverse the space of possible musical outcomes, ensuring that a system can produce variations rather than the same output every time.

\subsection{Structured vs. Unstructured Randomness}

A crucial distinction is whether randomness is applied in an unstructured way or is guided by structure. Unstructured randomness refers to unguided chance – for example, picking pitches completely at random or generating audio waveforms from noise without constraints. In contrast, structured uncertainty involves randomness within an organised framework. A structured approach might use a probabilistic model that has learned musical patterns, so that its random outputs still follow the statistical tendencies of a genre. For example, an improvisation algorithm might randomly select the next note but weighted by a learned scale or harmony, as opposed to any of the 12 tones with equal probability. By embedding randomness inside a learned model or set of rules, the system ensures that even unpredictable elements remain musically meaningful. This concept aligns with the observation that structured generative models can offer more controllability and interpretable variation than black-box ones. Figure~\ref{fig:structured-randomness} illustrates this distinction, showing how even minimal structural constraints (e.g., a Markov model) can yield musically coherent outputs compared to purely random generation.

\subsection{Human Control and Co-Creative Potential}

The degree of human involvement in steering a system also influences how randomness is used. Some AI music systems function autonomously: the user provides minimal input (perhaps just a prompt or a “generate” command) and the system’s internal randomness drives the output. Other systems invite the user into the loop, allowing interactive influence on the generative process. For example, an artist might adjust a temperature knob to make an RNN-based melody generator more or less random, or use a tool like the Wekinator to map live gestures to sound parameters in real time. 

Interactive systems let users modulate the amount of randomness and impose their own structure moment-to-moment. This can turn potentially chaotic outputs into something intentionally expressive, as the human guides the algorithm. Real-time co-improvisation systems take this further: they adapt to a musician’s input on the fly, using randomness to respond in novel ways while remaining in context. Thus, an important aspect of designing for randomness is deciding who (or what) controls the uncertainty – the algorithm alone, the user, or a combination of both.


From a co-creativity perspective, these design choices are significant. Prior work suggests that systems best support co-creative outcomes when they enable mutual influence, contextual responsiveness, and shared agency \citep{lubart2005can, davis2016empirically}. Co-creative potential increases when randomness is embedded within structured, real-time, human-guidable interactions. In such contexts, unpredictability becomes a source of inspiration rather than disruption—supporting emergent meaning while maintaining coherence. This balance of unpredictability and control is critical to fostering improvisational dynamics and sustaining a sense of shared authorship. 

These dynamics resonate with the theory of participatory sense-making, which describes how creativity can arise through reciprocal interaction between autonomous agents \citep{de2007participatory}. When uncertainty is scaffolded by temporal structure, feedback channels, and mutual responsiveness, it invites exploration without overwhelming the agent. Rezwana and Maher’s COFI framework further underscores this point, identifying improvisational systems with parallel participation and spontaneous initiative as particularly fertile for emergent creativity, while noting that a lack of structured interaction (e.g., communication or shared rhythm) can impair co-creative flow \citep{rezwana2023designing}. Structured uncertainty, shaped through active human engagement and situated interaction design, can thus serve not only musical novelty but also shared creative intent.

\section{Themes in Handling Randomness in AI Music Systems}

Several broad themes emerge when comparing how different AI music systems incorporate randomness and uncertainty. The following four key dimensions are laid out:

\subsection{Randomness Mechanisms and Control Parameters}

Different systems leverage different mechanisms for randomness, each implying distinct creative possibilities based on how unpredictability interacts with human perception and action. Sequence models like Melody RNN use probabilistic sampling to determine each successive element, often exposing a temperature parameter that modulates the entropy of predictions: higher temperatures yield more surprising and divergent outputs, while lower temperatures produce safer, more predictable continuations. In contrast, GAN-based systems such as Musika draw a random latent vector once to generate an entire passage; techniques like the truncation trick can then constrain the variability of this vector to reduce output wildness. These differences reflect varying approaches to what \citet{beghetto2021there} calls structured uncertainty—a balance of openness and constraint that invites creative responses to unpredictability. In this context, the structure of randomness becomes an interface for creativity, shaping how users engage with surprise, doubt, and possibility.

VAE-based models (e.g., RAVE) sample from a latent space where each dimension has learned significance, so a user might navigate that space to invoke randomness along certain musical dimensions (like timbre variation) while keeping others constant. Hierarchical models like MIDI-DDSP introduce randomness at multiple levels (performance and timbre) but in a controlled way – if the user does not specify certain details, the model’s priors will randomly fill them in with plausible values. Each mechanism comes with control levers: sampling strategies, latent space interpolations, or user interfaces that adjust how unpredictable the generation is.

\subsection{Degree of Structure in Random Outputs}

A central concern in the design of generative music systems is how randomness is constrained by musical structure. Structured uncertainty—randomness shaped by learned or explicitly designed constraints—has been described as beneficial for creative outcomes in both educational and collaborative domains \citep{beghetto2019structured, ibert2018uncertainty}. While these frameworks are not specific to music, they offer a useful lens for interpreting how AI systems manage uncertainty in musical contexts.

Some systems permit largely unconstrained output and rely on statistical regularities in training data to yield musically coherent results. For instance, Musika’s GAN-based architecture is mostly unconditional; it relies on the model’s latent space and the training corpus to ensure stylistic plausibility. The addition of a “global context” mechanism in Musika introduces a persistent conditioning signal that promotes stylistic continuity—a way of structuring an otherwise open-ended generative process.

In contrast, systems like Somax 2 impose strict structural constraints on their randomness. Somax 2’s improvisation process is limited to recombining fragments from a known corpus, meaning its stochastic choices always remain within the stylistic and material boundaries of the source. This ensures a high degree of musical coherence, particularly important in real-time co-improvisation scenarios.

Sequence models such as Melody RNN occupy a middle ground. Their outputs are shaped by learned temporal and tonal relationships—if trained effectively, they tend not to jump erratically across unrelated keys—but they lack external rule enforcement unless explicitly conditioned.

These examples illustrate a continuum of generative freedom, ranging from open-ended to highly constrained. Designers must navigate this trade-off: too few constraints risk incoherence, while too many may limit creative potential. Finding productive zones of structured uncertainty remains a key challenge in crafting co-creative musical tools.


\subsection{Real-Time Interactivity and Adaptation}

Whether a system operates in real time or offline significantly shapes how randomness is employed. Real-time systems—such as RAVE, Wekinator, and Somax 2—must generate outputs rapidly and often in response to unpredictable human or environmental inputs. As a result, their generative processes must be both efficient and reliably constrained: there is little opportunity for extensive computation or for filtering out undesirable outputs. RAVE, for example, operates at 20× real-time speed on a CPU, enabling low-latency live performance; its random variations unfold fluidly as the performer manipulates parameters. Somax 2 listens continuously to live musicians and adapts its behavior accordingly, using stochastic processes to introduce variation while remaining in synchrony with the ongoing performance.

By contrast, offline systems such as Melody RNN or Musika can generate multiple candidate outputs and allow the user to curate the results. These systems can accommodate more exploratory or unstructured randomness, as there is time for post-hoc selection or refinement. Real-time use, however, demands that randomness behave responsibly—surprises are welcome, but only if they cohere musically in the moment.

In such contexts, structured uncertainty becomes particularly valuable. Purely random outputs risk disrupting the flow of interaction, whereas context-sensitive variation can support emergent musical structures and invite creative response. This orientation aligns with the ethos of ‘Live Algorithms’—autonomous systems designed to improvise alongside human performers—where real-time adaptability and meaningful interaction are central \citep{blackwell2012live, knotts2020live, mackey2012musical}.

\subsection{Co-Creative Roles of Randomness}

Lastly, the role of the human user defines how randomness is employed. Autonomous composition systems (e.g., Musika or a high-temperature Melody RNN) may generate a lot of material independently, positioning the human primarily as curator or editor. By contrast, assistive generation tools (like MIDI-DDSP) and interactive systems (like Wekinator) afford greater human control. MIDI-DDSP lets users intervene at the note, performance, or timbral level, choosing where randomness adds to expression (e.g., via humanised microtiming) and where to fix decisions. Wekinator allows musicians to define mappings that control randomness in real time, effectively making the user a co-designer of the system’s generative behavior. This aligns with the view of machine learning as a meta-instrument—flexible, expressive, and shaped through interaction \citep{fiebrink2017machine}.

Somax 2 positions the system as an improvising partner, responding dynamically to live input with stochastic variation. Here, randomness becomes part of a reciprocal, participatory exchange. Co-creative systems that allow both the human and the AI to adapt in real time support what \citet{rezwana2023designing} call interactional sense-making—the mutual shaping of creative meaning through collaborative dynamics.

This collaborative dynamic also resonates with \citet{davis2017creative} creative sense-making framework, which affirms the emergent, time-sensitive nature of improvisational co-creation. Their work shows that meaningful co-creativity relies not just on novelty but on the evolving interplay of clamped and unclamped cognitive states—moments where collaborators settle into or reconfigure their shared understanding. When randomness is situated within this interactive loop—providing surprise without destabilising the process—it can complement rather than derail co-creation. \citet{lubart2005can} similarly argued that systems should allow for varying degrees of autonomy and intervention, adapting to user goals across different phases of the creative process.

Ultimately, the co-creative potential of an AI system hinges on how it balances generativity and steerability. Systems that allow users to guide or respond to randomness in real time tend to support richer collaboration than those where randomness occurs in isolation. This reflects broader notions of reciprocal adaptation and creative responsiveness in human-AI interaction \citep{rezwana2023designing, davis2016empirically, lubart2005can}, asserting that randomness is most effective when it contributes to a dynamic, emergent partnership rather than a one-sided generation pipeline.

Having outlined these themes, I now turn to six specific AI music systems that exemplify different approaches. In the next section, these systems are compared side by side, detailing their design and how they handle randomness and uncertainty.

\section{Randomness and Co-Creative Potential: A Comparative Overview of AI Music Systems}

To ground the discussion, a comparative review of six selected AI music systems is presented, each illustrating distinct ways of handling randomness and uncertainty. While these systems differ significantly in scope, modality, and intended use, they are included here for their illustrative value in showing how randomness, structure, and control are handled across a representative spectrum. This diversity is a limitation if aiming for uniform comparison, but it strengthens the broader argument by showcasing how different design decisions shape the co-creative affordances of AI music tools.

The system selection was based on two main criteria. First, the systems represent diversity by design, as reflected in the spectrum of input and output modalities they employ (e.g., Somax2: mixed modality; Melody RNN: symbolic; Musika: audio). Second, the models are open-source and relatively influential, meaning that they are well-known within the field and readily available for experimentation and use by practicing artists.

Table 1 provides an overview of key characteristics for each system. I then describe each system’s approach in more detail, stressing how they balance stochastic creativity with musical structure and user control. Note the columns \textit{Randomness Type} and \textit{Level of Human Control}: while these categories are not necessarily mutually exclusive in terms of how Structured Uncertainty is framed in the paper (e.g., a relatively unstructured level of randomness tends to bear at least some relationship to the level or perception of human control), they have nonetheless been organised using the following criteria. 

\textit{Randomness Type} is characterised primarily by the degree to which constraints are imposed by model design. For example, Musika is relatively unstructured because it autonomously generates novel music within broad confines based on its training data. This partially relates to the low \textit{Level of Human Control}, but the limited number of readily exposed parameters available to artists in performance also contributes significantly to this, functioning as a constraint in its own right. Conversely, Somax2 employs randomness, but it is much more context-sensitive by design, drawing directly from and recombining source material. Again, this influences \textit{Level of Human Control}, but the interface and interactive performance design of the tool also have a significant bearing on this.

\begin{table}[ht]
  \caption{Comparative Features of Selected AI Music Systems in Handling Uncertainty.}
  \label{tab:system-summary}
  \centering
  \begin{tabular}{lllll}
    \toprule
    \textbf{System} &  \textbf{Real-Time} & \makecell{\textbf{Randomness} \\ \textbf{Mechanism}} & \makecell{\textbf{Randomness} \\ \textbf{Type}} & \makecell{\textbf{Level of} \\ \textbf{Human Control}} \\
    \midrule
    Musika                 & No   & Random latent vectors     & Unstructured     & Low \\
    Melody RNN          & No   & Sampling temperature      & Semi-structured  & Medium \\
    Midi-DDSP             & No   & Hier. sampling (perf./timbre)     & Structured       & High \\
    RAVE                   & Yes  & Sampling from VAE latent space      & Mixed            & Medium-High \\
    Wekinator        & Yes  & Depends on input data     & Mixed            & High \\
    Somax 2            & Yes  & Contextual prediction     & Structured       & High \\
    \bottomrule
  \end{tabular}
\end{table}

\subsection{Musika}

Musika is a deep learning system for raw audio music generation. It employs a GAN (Generative Adversarial Network) architecture to produce infinite-length waveform music, with randomness introduced through the sampling of latent vectors for the GAN’s generator. Unlike many audio GANs that produce fixed-length clips, Musika uses a latent coordinate system that arranges latent inputs along a path to ensure continuity between segments. A global context vector further conditions the output, promoting stylistic coherence over time. These features impose structure on an otherwise unstructured process, guiding generation to avoid abrupt changes and maintain a unified musical style.




Musika’s output remains largely unconditional and unguided by external scores or real-time input once trained—it exemplifies a system where randomness generates novel music within the broad confines of a learned genre. It is not designed for real-time interaction; generation is offline, and user control is minimal (aside from model selection or training).

Its co-creative potential in interactive settings is thus low. Musika functions more as an autonomous generator for background music or inspiration. Nevertheless, it demonstrates that even with unstructured randomness (i.e., GAN latent space sampling), careful model design can yield outputs that sound coherent and musically interesting. Co-creativity remains minimal, with Musika acting more as a content generator than a responsive partner \citep{pasini2022musika}.

\subsection{MIDI-DDSP}

MIDI-DDSP is a hierarchical generative system that bridges symbolic music representation and neural audio synthesis. It takes a MIDI note sequence as input and produces realistic instrument audio using Differentiable Digital Signal Processing (DDSP) techniques. The system emphasises fine-grained control and interpretability, decomposing generation into three stages—notes, performance, and synthesis. The user provides the notes (e.g., a score or melody); the model then generates performance parameters (such as dynamics, vibrato, articulation), and finally synthesises the audio using a neural oscillator.

Randomness arises primarily when the system autonomously fills in performance or timbral details via its learned prior models. Given a melody, for instance, it can sample likely expressive timing or intensity patterns, yielding a plausible performance with slight variation. 

Because each level is interpretable, a user can override randomness—for example, by specifying their own dynamics—or let the model’s stochastic suggestions stand. MIDI-DDSP operates offline (rendering audio files rather than responding live) but serves as an assistive composition tool.

A musician can iterate with it: providing a score, reviewing different performative renditions, and selecting or tweaking results. Randomness is structured by the model’s hierarchy: it will not invent new notes (unless instructed), and its expressive variations reflect real performances, making them realistic rather than arbitrary. Human control is high, and co-creativity is at a medium level—the system does not generate full compositions, but acts as a creative assistant offering nuanced, interpretable variation. The separation of musical parameters allows users to guide or override randomness, supporting a collaborative workflow where expressive details emerge through interplay \citep{wu2021midi}.

\subsection{Melody RNN}

Melody RNN is one of the early deep-learning-driven music generators from Google’s Magenta project. It uses a recurrent neural network (LSTM) to generate monophonic melodies, one note at a time. Trained on a large corpus, the model learns typical melodic sequences and patterns. During generation, it introduces randomness by sampling the next note from the LSTM’s output distribution. As a result, the same initial melody (or “primer”) can produce different continuations on different runs, especially at higher sampling temperatures.

This randomness is moderately structured: since probabilities are learned from real music, the model’s outputs tend to follow musical logic (e.g., staying in key or repeating rhythmic motifs). However, it lacks explicit global structure beyond the RNN’s memory. Melody RNN is not a real-time interactive system; it generates melodies offline in seconds. User control is limited to selecting the model variant (Basic, Lookback, or Attention RNN) and adjusting parameters like temperature or melody length. In practice, composers may generate a batch of melodies and select or develop promising ones.

Its co-creative potential is relatively low in live contexts, but as an idea generator, it is effective for sparking inspiration. Melody RNN demonstrates that even a simple sequence model with randomness can produce musically coherent results—illustrating the power of learned structure. Without explicit music theory, it often yields repetition and development drawn from training data. While it does not support live co-creation, its capacity to generate varied melodic ideas makes it a useful prompt for composers.\footnote{\url{https://magenta.tensorflow.org/2016/07/15/lookback-rnn-attention-rnn/}}

\subsection{RAVE}

RAVE (Realtime Audio Variational autoEncoder) is a neural model for high-quality audio synthesis, notable for its real-time capabilities. It uses a variational autoencoder architecture: an encoder compresses audio into a low-dimensional latent code, and a decoder reconstructs audio from it. Randomness enters through sampling or manipulating the latent space.

Because the VAE includes a stochastic component (the encoder outputs a distribution), users can sample random latent vectors to generate new sounds. RAVE’s authors introduced a two-stage training process (reconstruction followed by adversarial fine-tuning) and post-training analysis to control the trade-off between fidelity and variability. Users can thus tune the system to be more precise or more generative, depending on the creative need.

Critically, RAVE runs fast enough for interactive use—generating 48 kHz audio significantly faster than real time. Musicians have integrated it into performance setups, using MIDI or other controllers to explore the latent space or morph between timbres on the fly. In these contexts, RAVE’s randomness is often user-guided: performers might randomise certain dimensions or inject noise to produce glitch effects, while stabilising others (e.g., pitch contour via MIDI).

Co-creatively, RAVE acts as a flexible instrument—the user explores, and the model offers a universe of variations beyond what a hand-designed synthesiser might easily yield. Its speed and interactivity make it well-suited to co-creative performance \citep{caillon2021rave}.

\subsection{Wekinator}

While Wekinator is not a generative model in the traditional sense, its inclusion highlights how artists can construct real-time mappings that introduce or modulate randomness. Wekinator is an interactive machine learning toolkit rather than a fixed music generator. Developed by Rebecca Fiebrink, it enables musicians and artists to build real-time mapping systems by training supervised learning models on the fly. In a typical use case, a user connects inputs (e.g., sensor data) to outputs (e.g., synth parameters), provides training examples, and Wekinator learns a model (e.g., neural network or regression) that maps inputs to outputs in real time.

Randomness in Wekinator is not inherent—if the user provides deterministic examples, outputs will be deterministic. However, users can deliberately incorporate randomness into the mapping design. For instance, a noise signal could be used as an input to generate changing outputs, or inconsistent training examples might produce nuanced, unpredictable responses. Wekinator hands full control of structure versus randomness to the user. It can produce precise mappings or chaotic ones, depending on intent.

Wekinator is fully real-time and encourages live experimentation. Human control is maximal—the user defines the entire mapping logic. Its co-creative potential is also high, though it differs from generative systems: Wekinator does not generate music independently, but enables interactive instruments where creativity emerges from human–machine interplay. Wekinator exemplifies a philosophy of engaging with randomness through direct human mediation: neutral in itself, it empowers artists to shape indeterminacy in performance \citep{fiebrink2010wekinator}.

\begin{figure}[htbp]
  \centering
  \includegraphics[width=0.7\linewidth]{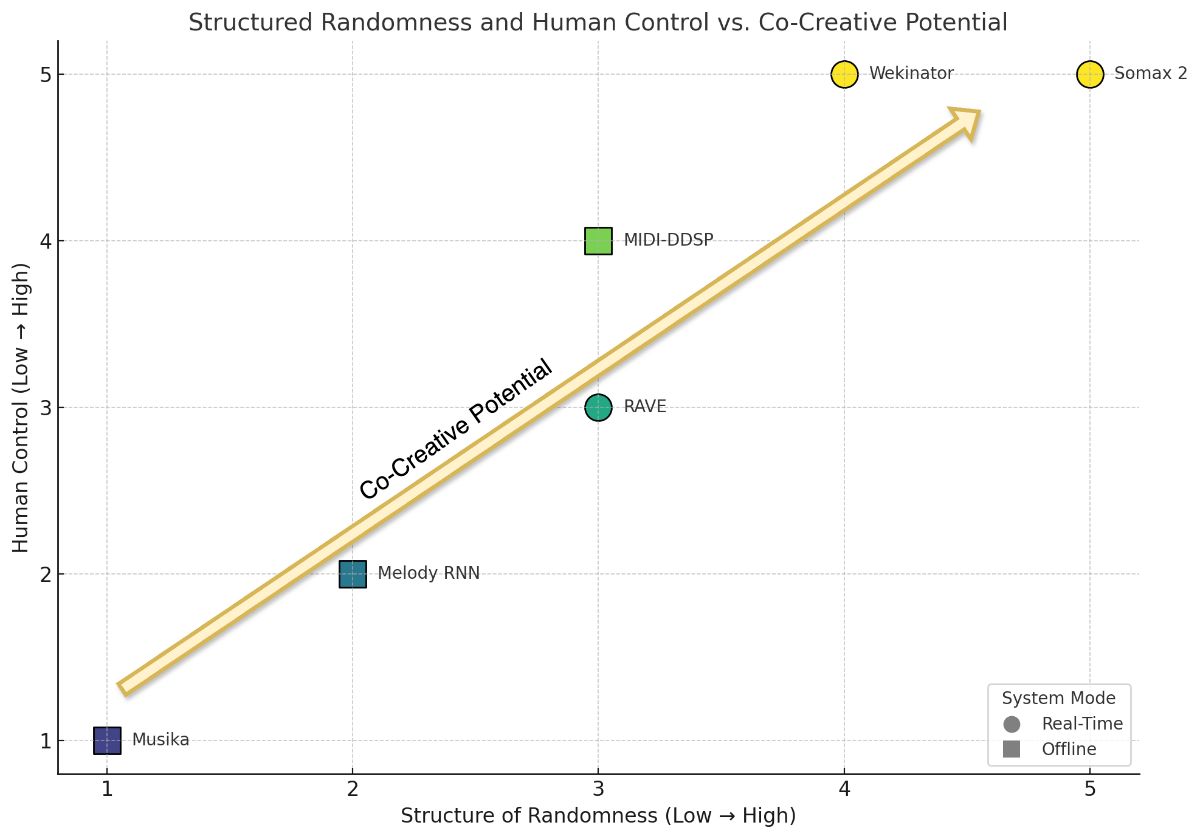}
  \caption{
  A conceptual map of co-creative potential across six AI music systems. Each system is plotted according to the degree of human control (vertical axis), the structure of randomness (horizontal axis), and real-time capability (shape). Systems higher and further to the right—those with structured uncertainty, strong human guidance, and real-time interactivity—tend to exhibit greater co-creative potential.
  }
\end{figure}


\subsection{Somax 2}

Somax 2 is a state-of-the-art system for machine co-improvisation with human musicians. It builds a statistical model (a corpus-based memory) from musical material provided to it (e.g., recordings or MIDI files of a particular style or artist). As music plays—whether from a human improviser or pre-recorded streams—Somax’s machine listening modules analyse it and update the model in real time.

The result is that Somax 2’s output remains stylistically consistent with its source material; it does not generate novel material ex nihilo, but recombines and transforms what it has learned. Randomness enters in selecting which learned fragment to play next, or how to vary it. This randomness is highly context-sensitive: the system ensures that any choice aligns with the live input’s harmony, timing, and style. In this sense, Somax 2 practices structured improvisation—using statistical modeling to create a memory structure it can navigate to form new, stylistically coherent musical sequences.

Because Somax 2 operates live, it must react quickly; its generative algorithms are fast and incremental, producing output in sync with human musicians. Users interact primarily by preparing the training corpus and adjusting high-level behavioral parameters (e.g., how adventurous the agents should be). During performance, interaction is embodied—Somax listens and adapts, incorporating new motifs introduced by the human performer, if they relate to its learned corpus. The performer experiences Somax as a semi-autonomous partner: it contributes its own creative flavor while remaining grounded in the musical language defined by the user. Somax 2 exemplifies real-time co-creativity, harnessing randomness to generate stylistically coherent, responsive improvisation \citep{borg2019somax}.

\section{Discussion and Conclusion}

This review reveals a spectrum of strategies for balancing randomness and structure in AI music systems, each shaped by its creative context. Autonomous models like Musika rely on latent space exploration for novelty, while systems like Wekinator and Somax 2 foreground human input and real-time responsiveness. Between these lie hybrid approaches (e.g., RAVE, MIDI-DDSP) that embed randomness within controllable frameworks.

A key insight is that randomness alone does not guarantee creative value—it must be contextualised. Offline tools can afford greater unpredictability, as users curate results. In contrast, live or interactive systems demand structured uncertainty that aligns with human intent and timing. This supports not only musical coherence, but also richer co-creative potential: systems like Somax 2 and Wekinator demonstrate how real-time, guided randomness can contribute to responsive and reciprocal creativity.



The core contribution of this paper is a thematic review of how AI music systems engage with randomness and uncertainty—identifying design patterns that support coherence, control, and co-creativity. By comparing six diverse systems, practical strategies for embedding stochastic inspiration within meaningful interaction frameworks are surfaced. I suggest that co-creative potential is highest when randomness is harnessed through real-time systems that afford user guidance and embed structured uncertainty—where randomness is not eliminated, but shaped by learned patterns, constraints, or human input. This aligns with participatory sense-making theories in collaborative creativity. \textit{To my knowledge, this is the first thematic review to examine randomness in AI music systems through the lens of structured uncertainty and co-creative affordances}. 

I acknowledge that the selection of systems, while representative, is by no means exhaustive. The diversity of design goals, modalities, and user contexts across AI music tools makes systematic comparison challenging. Nonetheless, this thematic lens surfaces useful patterns that may inform future design and evaluation strategies. 

To address the practical implications of this review, here are some clear takeaways that developers and artists can carry forward when designing co-creative AI music tools:

\begin{itemize}
    \item \textbf{Leverage structured uncertainty as part of the design philosophy:} Draw explicitly from the crucial interplay between the type and structure of randomness and the level of human control, using these insights to steer the co-creative potential of a system. Model and system selection should consider how these factors interact to support purposeful exploration while maintaining musical coherence.
    
    \item \textbf{Develop formalised benchmarking within evaluation frameworks:} Incorporate structured uncertainty explicitly as part of creative system evaluation, assessing how uncertainty is applied, how effective it is in practice, and how it relates to the efficacy of a system for artistic use. This could help designers identify the utility of structured uncertainty in their tools.
    
    \item \textbf{Prioritise human-centric design:} Use structured uncertainty, which draws heavily on established human creativity research, as tangible impetus and justification for system utility, asking “Is this human-centred by design?” as a checkpoint to ensure the artist remains actively engaged in the creative loop. This orientation supports systems that are not only technically capable but also meaningful and empowering for artistic practice.
\end{itemize}

Together, these practical considerations reinforce the central claim that structured uncertainty, when deliberately embedded within human-centred and responsive system design, can bolster the co-creative potential of AI music tools while supporting artistic coherence and meaningful exploration.

\begin{ack}
This publication has emanated from research conducted with the financial support of Taighde Éireann – Research Ireland under Grant number 18/CRT/6222. For the purpose of Open Access, the author has applied a CC BY public copyright licence to any Author Accepted Manuscript version arising from this submission.
\end{ack}


\pagebreak

\bibliographystyle{apalike}   
\bibliography{references}  

\end{document}